\documentclass[10pt,conference]{IEEEtran}
\newlength{\pic}
\newlength{\bigPic}
\usepackage{graphicx}
\usepackage{amsthm,amssymb}
\usepackage{pgf,pgfarrows,pgfnodes}
\usepackage{tikz}
\usepackage{algorithm, algorithmic}
\usepackage{lpic}
\usetikzlibrary{arrows,decorations.pathmorphing,backgrounds,positioning,fit,calc} 
\newtheorem{lemma}{Lemma}
\newtheorem{theorem}{Theorem}

\theoremstyle{definition}
\newtheorem{definition}{Definition}

\theoremstyle{remark}

\theoremstyle{plain}

\newcommand{\cov}{\mathop{\mathrm{cov}}}
\newcommand{\var}{\mathop{\mathrm{var}}}

\begin{document}

\title{Physical-Layer Security over Correlated\\ Erasure Channels}

\author{\IEEEauthorblockN{Willie K. Harrison\IEEEauthorrefmark{1},
Jo\~{a}o Almeida\IEEEauthorrefmark{2},
Steven W. McLaughlin\IEEEauthorrefmark{1} and
Jo\~{a}o Barros\IEEEauthorrefmark{2}}

\IEEEauthorblockA{\IEEEauthorrefmark{1}School of Electrical and Computer Engineering,
Georgia Institute of Technology,
Atlanta, Georgia 30332\\ Email: \{harrison.willie@, steven.mclaughlin@provost.\}gatech.edu}
\IEEEauthorblockA{\IEEEauthorrefmark{2}Instituto de Telecomunica\c{c}\~{o}es, Departamento de Engenharia Electrot\'{e}cnica e de Computadores\\
Faculdade de Engenharia da Universidade do Porto, Portugal,
Email: \{jpa, jbarros\}@fe.up.pt}
}

\maketitle

\renewcommand{\thefootnote}{}
\footnotetext{The research in this paper was partially funded by the US National Science Foundation (Grant NSF--0634952), the Luso-American Foundation (FLAD), the Funda\c{c}\~{a}o para a Ci\^{e}ncia e Tecnologia (FCT Scholarship SFRH/BD/60831/2009), and the WITS project (Grant PTDC/EIA/71362/2006).
}

\renewcommand{\thefootnote}{\arabic{footnote}}

\begin{abstract}

We explore the additional security obtained by noise at the physical layer in a wiretap channel model setting. Security enhancements at the physical layer have been proposed recently using a secrecy metric based on the degrees of freedom that an attacker has with respect to the sent ciphertext.  Prior work focused on cases in which the wiretap channel could be modeled as statistically independent packet erasure channels for the legitimate receiver and an eavesdropper. In this paper, we go beyond the state-of-the-art by addressing correlated erasure events across the two communication channels. The resulting security enhancement is presented as a function of the correlation coefficient and the erasure probabilities for both channels. It is shown that security improvements are achievable by means of judicious physical-layer design even when the eavesdropper has a better channel than the legitimate receiver. The only case in which this assertion may not hold is when erasures are highly correlated across channels. However, we are able to prove that correlation cannot nullify the expected security enhancement if the channel quality of the legitimate receiver is strictly better than that of the eavesdropper.

\end{abstract}

\section{Introduction}\label{sec:intro}
The origins of physical-layer security were laid down by Shannon \cite{Shannon49} and Wyner \cite{Wyner75} in seminal papers. Due to these works, and the contributions of others, it is known that channel codes exist that can achieve both \emph{reliability} and \emph{security} in relevant communication scenarios. The wiretap channel model presented in \cite{Wyner75} is instrumental to this fact. Since that time, the goals of physical-layer security research have been twofold: first, to develop understanding of theoretical fundamental bounds on secrecy; and second, to design practical codes which achieve the secrecy bounds (see \cite{BlochBook} and its references). Unfortunately, meeting theoretically-achievable secrecy bounds within a practical coding scheme often requires assumptions which may be unrealistic in practice. For example, some designs offer reliability to legitimate parties based on the premise that the main channel is noiseless \cite{Thangaraj07, Arun10_ITW}. Other designs assume that the encoder has perfect channel state information (CSI) for friendly parties and malicious parties alike \cite{Mahdavifar10}. Finally, nearly every physical-layer security scheme cannot maintain secrecy when an eavesdropper has a consistent advantage in channel quality.

As a result, the point was made in \cite{Harrison10_TIFS} that physical-layer security should be thought of as an extra layer of secrecy which can aid security efforts at other layers in the communications protocol stack. The coding scheme presented in \cite{Harrison10_TIFS,Harrison10_ITW} was developed with the intent of avoiding unrealistic assumptions for practical application of a secrecy code which enhances security using stopping sets in punctured low-density parity-check (LDPC) codes. The added security was measured in the number of degrees of freedom $D$ in an eavesdropper's information regarding the ciphertext, or equivalently in equivocation \cite{Harrison10_TIFS}. The design is impervious to a lack of CSI, because it ensures that $\mathbb{E}[D]$ exceeds any fixed $\beta$ for nearly every possible combination of channel parameters as the dimension of the encoder increases. Thus, security enhancement occurs in the system even when an eavesdropper has a better channel than the legitimate receiver.

Despite the advantages of this scheme, as with other practical designs, a simplifying assumption was made in the security analysis where erasure events for legitimate receivers and eavesdroppers were assumed to be statistically independent. However, in real radio environments, channels from a transmitter to different receivers may be correlated depending on the physical deployment of the receiver antennas, the availability of line-of-sight, and the presence or absence of scatterers at the transmitter and receivers \cite{Lee73,Jeon08}.

Therefore, we address the effects of correlation between packet erasures at an intended receiver and packet erasures at the eavesdropper on the performance of the physical-layer security design from \cite{Harrison10_ITW}. The key contributions of this work are the following:
\begin{itemize}
  \item \emph{Security Analysis for Correlated Erasures:} Using degrees of freedom in the eavesdropper's knowledge of the ciphertext, the security analysis is made for the correlated packet erasure channel model. Comparisons are given between the correlated and uncorrelated cases.
  \item \emph{Correlation Coefficient Boundaries:} Bounds on the correlation coefficient are derived, and analysis of security enhancement is provided assuming best and worst correlation conditions.
  \item \emph{Security Enhancement:} It is shown that in many cases security enhancement can still be obtained, even when the eavesdropper has a better channel than the legitimate receiver and erasures are correlated. We also prove that correlation cannot reduce $\mathbb{E}[D]$ to zero if the legitimate receiver's channel quality is strictly better than that of the eavesdropper.
\end{itemize}

The rest of the paper proceeds as follows. In Section \ref{sec:modelAndCorrelation} we discuss the correlated wiretap channel model with feedback, and boundary properties of the correlation coefficient. We briefly discuss stopping sets, degrees of freedom as a security metric, and present the encoder and decoder from \cite{Harrison10_ITW} in Section \ref{sec:system}. The security results for the correlated wiretap packet erasure channel model are given in Section \ref{sec:decodeAndSecrecy}, and conclusions are provided in Section \ref{sec:conclusion}.

\section{Correlated Channel Model}\label{sec:modelAndCorrelation}

\begin{figure}
\begin{center}
  \begin{tikzpicture}
    [node distance=0.8cm, rounded corners=2pt, channel/.style={rectangle,draw=blue,fill=blue!10,thick,
    text centered,minimum size=8mm},
    boxedNode/.style={rectangle,draw,fill=black!10,thick,text centered, minimum size=8mm},
    inner sep=1mm]
    \node (Alice) {Alice};
    \node [boxedNode] (Encoder) [right=of Alice] {Encoder};
    \node [channel] (Qm) [right=of Encoder,label=above:$Q_m$] {PEC($\delta$)};
    \node [boxedNode] (Decoder) [right=of Qm] {Decoder};
    \node (Bob) [right=of Decoder] {Bob};
    \node [channel] (Qw) [below=of Qm,label=above:$Q_w$] {PEC($\epsilon$)};
    \node [boxedNode] (Decoder_w) [right=of Qw] {Decoder};
    \node (Eve) [right=of Decoder_w] {Eve};
    \draw[->] (Alice) -- node [above] {$M$} (Encoder);
    \draw[->] (Encoder) -- node [above] {$X$} (Qm);
    \draw[->] (Qm) -- node [above] {$Y$} (Decoder);
    \draw[->] (Decoder) -- node [above] {$\tilde{M}$} (Bob);
    \draw[->] (Qw) to node [above] {$Z$} (Decoder_w);
    \draw[->] (Decoder_w) to node [above] {$\hat{M}$} (Eve);
    \draw[->] ($(Encoder.east) + (4mm,0)$) |- (node cs:name=Qw,anchor=west);
    \draw [->] (Bob) -- ++(0,1.1)
        -| node[near start, above] {Feedback Channel} (Encoder);
  \end{tikzpicture}
\end{center}
\caption{Wiretap channel model with feedback assuming correlated packet erasures in the main channel $Q_m$ and the wiretap channel $Q_w$.}\label{fig:wiretapChannel}
\end{figure}
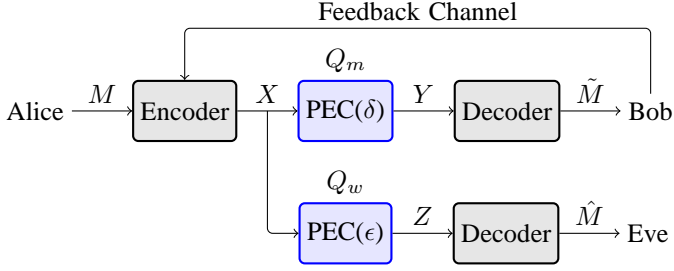

We consider the wiretap channel model with memoryless packet erasure channels (PEC), and add automatic repeat request (ARQ) for authenticated users as shown in Fig. \ref{fig:wiretapChannel}. A user named Alice encodes an encrypted and compressed message $M$ into a set of $\eta$ packets denoted $X$ according to a specific encoding rule. The packets are then transmitted over the \emph{main} channel $Q_m$ to an intended recipient named Bob who receives the set of packets $Y$ where packet erasures occur  with probability $\delta$. Bob is permitted to request the retransmission of any missing packets using an authenticated feedback channel. After obtaining all transmitted packets, he applies the decoding rule and obtains an estimate of the secret message $\tilde{M}$. An eavesdropper named Eve also observes the transmitted packets, albeit through a different channel called the \emph{wiretap} channel $Q_w$. Eve obtains the packets $Z$ through $Q_w$, by observing the initial transmission as well as any retransmitted packets. Erasures occur in this channel with probability $\epsilon$. She also attempts to decode the data and obtains an estimate of the secret message $\hat{M}$.

Since $Q_m$ and $Q_w$ are memoryless, erasures occur independently with respect to one another in a given channel; however, erasures across channels are correlated with correlation coefficient $\rho$. Let $E_m$ and $E_w$ be Bernoulli random variables which take on values in the set $\{0,1\}$, where one signifies erasure and zero signifies error-free reception of a packet. Thus, $\Pr(E_m=1) = \delta$ and $\Pr(E_w=1)=\epsilon$. The covariance of two random variables $A$ and $B$ is defined as $\cov(A,B)=\mathbb{E}[(A-\mathbb{E}[A])(B-\mathbb{E}[B])]$, and the variance of a random variable $A$ can be expressed as $\var(A) = \cov(A,A)$ \cite{Grimmett_ProbBook}. Given these definitions, the Pearson correlation coefficient between random variables $E_m$ and $E_w$ is  \cite{Grimmett_ProbBook}
\begin{eqnarray}\label{eq:rho}
  \rho &=& \frac{\cov{(E_m,E_w)}}{\sqrt{\var{(E_m)}\var{(E_w)}}} \nonumber \\
  &=& \frac{\mathbb{E}[E_mE_w]-\mathbb{E}[E_m]\mathbb{E}[E_w]}{\sqrt{(\mathbb{E}[E_m^2]-\mathbb{E}[E_m]^2)(\mathbb{E}[E_w^2]-\mathbb{E}[E_w]^2)}} \nonumber \\
  &=& \frac{\mathbb{E}[E_mE_w]-\delta\epsilon}{\sqrt{\delta(1-\delta)\epsilon(1-\epsilon)}}.
\end{eqnarray}
The last step is made using the first and second moments of a Bernoulli random variable, where $\mathbb{E}[E_m] = \mathbb{E}[E_m^2]=\delta$ and $\mathbb{E}[E_w] = \mathbb{E}[E_w^2]=\epsilon$. Let $p_{ij}=\Pr(E_m=i,E_w=j)$ \cite{Zhao08}. Then, $\delta=p_{10}+p_{11}=\mathbb{E}[E_m]$ and $\epsilon=p_{01}+p_{11}=\mathbb{E}[E_w]$. It is trivial to show that $\mathbb{E}[E_mE_w]$ is equal to $\Pr(E_m=1,E_w=1)$. Thus, (\ref{eq:rho}) can be expressed as
\begin{equation}\label{eq:rhoB}
 \rho = \frac{p_{11}-\delta\epsilon}{\sqrt{\delta(1-\delta)\epsilon(1-\epsilon)}}.
\end{equation}

The Pearson correlation coefficient is commonly used to indicate the degree to which two random variables are similar. Although $|\rho|\leq 1$, it is a common misconception that $\rho$ can take on any value from $-1$ to $+1$. In reality, there are bounds to the coefficient which are a function of the distribution of the random variables involved \cite{Shih92}. In our case, we have allowed $\delta$ and $\epsilon$ to take on any value in $[0,1]$. We also know that $\delta = p_{11} + p_{10}$, $\epsilon = p_{11} + p_{01}$, and $p_{00} + p_{01} + p_{10} + p_{11} = 1$. Since $p_{ij}\geq0$ for $i,j\in\{0,1\}$, then
\begin{equation}\label{eq:p11Bounds}
 \max(\delta+\epsilon-1,0) \leq p_{11} \leq \min(\delta,\epsilon).
\end{equation}
The bounds on $p_{11}$ can be given as bounds on $\rho$ as
\begin{equation}\label{eq:rhoBounds}
 \frac{\max(\delta+\epsilon-1,0)-\delta\epsilon}{\sqrt{\delta\epsilon(1-\delta)(1-\epsilon)}}\leq \rho \leq \frac{\min(\delta,\epsilon)-\delta\epsilon}{\sqrt{\delta\epsilon(1-\delta)(1-\epsilon)}}.
\end{equation}
For example, if $\delta = 0.3$ and $\epsilon = 0.15$, then $-0.275\leq\rho\leq0.642$.

\section{Encoder and Decoder Design}\label{sec:system}

\begin{figure}
\begin{center}
  \begin{tikzpicture}
  [node distance=0.6cm, rounded corners=2pt, boxedNode/.style={rectangle,draw,fill=black!10,thick,
  text centered, minimum size=8mm},
  multiLine/.style={rectangle,text centered, text width=1.3cm, minimum size=8mm},
  boxMultiLine/.style={rectangle,draw,fill=black!10,thick,
  text centered, text width=1.3cm, minimum size=8mm},
  inner sep=1mm]
    \node [boxMultiLine] (Encoder) {LDPC Encoder};
    \node [boxMultiLine] (Puncture)  [right=of Encoder]  {Puncture Block};
    \node [boxedNode] (Buffer)  [right=of Puncture]  {Buffer};
    \node [boxedNode] (Interleaver)  [right=of Buffer]  {$\Pi$};
    \draw[->] ($(Encoder.west) + (-6mm,0)$) -- node [midway, above] {$M$} node [midway, below=14pt, text width=1.5cm, text centered] {$L$ blocks length $k$} (Encoder);
    \draw[->] (Encoder) to node [above, midway] {$B$} node [below=14pt, text width=1.5cm, text centered, midway] {$L$ blocks length $N$} (Puncture);
    \draw[->] (Puncture) to node [above] {$P$} node [below=14pt, text width=1.5cm, text centered, midway] {$L$ blocks length $n$} (Buffer);
    \draw[->] (Buffer) to (Interleaver);
    \draw[->] (Interleaver) to node [midway, above] {$X$} node [midway, below=12pt, text width=1.6cm, text centered] {$\eta$ packets size $\alpha L$} ($(Interleaver.east) + (6mm,0)$);
  \end{tikzpicture}
\end{center}
\caption{Detailed block diagram of the encoder. Number and size of blocks or packets are indicated at each step.}\label{fig:encoder}
\end{figure}
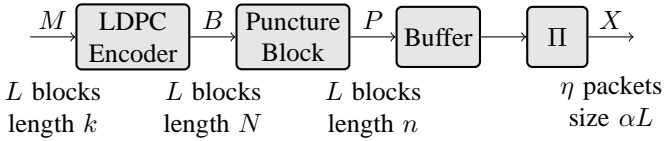

\begin{figure}
\begin{center}
  \begin{tikzpicture}
  [node distance=0.6cm, rounded corners=2pt, boxedNode/.style={rectangle,draw,fill=black!10,thick,
  text centered, minimum size=8mm},
  boxMultiLine/.style={rectangle,draw,fill=black!10,thick,
  text centered, text width=1.3cm, minimum size=8mm},
  inner sep=1mm]
    \node [boxedNode] (Buffer) {Buffer};
    \node [boxedNode] (Deinterleaver)  [right=of Buffer]  {$\Pi^{-1}$};
    \node [boxMultiLine] (messPass)  [right=of Deinterleaver]  {Message Passing};
    \node [rectangle,draw,fill=black!10,thick,text centered, text width=1cm, minimum size=8mm] (map)  [right=of messPass]  {Map to $\mathcal{M}$};
    \draw[->] ($(Buffer.west) + (-6mm,0)$) -- node [midway, above] {$Y$} node [midway, below=12pt, text width=1.6cm, text centered] {$\eta$ packets size $\alpha L$} (Buffer);
    \draw[->] (Buffer) to (Deinterleaver);
    \draw[->] (Deinterleaver) to node [above] {$\tilde{P}$} node [below=14pt, text width=1.5cm, text centered, midway] {$L$ blocks length $n$} (messPass);
    \draw[->] (messPass) to node [above] {$\tilde{B}$} node [below=14pt, text width=1.5cm, text centered, midway] {$L$ blocks length $N$} (map);
    \draw[->] (map) to node [midway, above] {$\tilde{M}$} node [midway, below=14pt, text width=1.5cm, text centered] {$L$  blocks length $k$} ($(map.east) + (6mm,0)$); 
  \end{tikzpicture}
\end{center}
\caption{Detailed block diagram of Bob's decoder. Number and size of blocks or packets are indicated at each step.}\label{fig:decoder}
\end{figure}
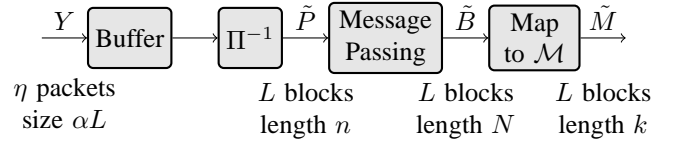

The goal of the security sub-system from \cite{Harrison10_ITW} is to inflict Eve with stopping sets in her received codewords, thus forcing her to guess the values of $D$ bits in the decoder. Even a single error in the guess is magnified so the ciphertext has bit-error rate 0.5. Block diagrams of the encoder and decoder are given in Figs. \ref{fig:encoder} and \ref{fig:decoder}. We now provide a brief explanation of the system.\footnote{For further details, the reader is referred to \cite{Harrison10_TIFS} and \cite{Harrison10_ITW}.}

\subsection{Stopping Sets}

Consider the message-passing (MP) decoder of an LDPC code $C$ \cite{Urbanke01} over the binary erasure channel (BEC). It helps to think of the $N-k\times N$ parity check matrix $H$, along with its Tanner graph representation. Let $V=(v_1,v_2,\ldots,v_N)$ be a set of vertices called \emph{variable nodes}. Also let $U=(u_1,u_2,\ldots,u_{N-k})$ be a set of vertices called \emph{check nodes}. Finally let $G_C$ be the Tanner graph representation of $H$, where $G_C$ is bipartite with bipartitions $V$ and $U$. An edge connects $u_i$ with $v_j$ if and only if $H_{i,j}=1$. Then all variable nodes connected to a particular check node form a checksum which is satisfied for all valid codewords in $C$.
\begin{definition}[Di, et. al. \cite{Di02StoppingSets}]\label{def:stoppingSet}
 A \emph{stopping set} is a set $\Lambda \subseteq V$ such that all check nodes in $N(\Lambda)$ are connected to $\Lambda$ by at least two edges, where $N(\Lambda)$ signifies the \emph{neighborhood} of $\Lambda$ and is defined as the set of all adjacent nodes to $\Lambda$ in $G_C$.
\end{definition}
Since by definition, the empty set is also a stopping set, each set of variable nodes has a maximal stopping set in it. If the standard MP decoder over the BEC is applied to a codeword where bits corresponding to $V'\subseteq V$ are erased, then the set of unknown bits after decoding is the maximal stopping set in $V'$ \cite{Di02StoppingSets}.

\subsection{Encoder}\label{subsec:encoder}

An encrypted and compressed message $M$ is grouped into $L$ blocks of length $k$. Allow the final block to be completed with random bits if necessary. Specifically, $M=(m^1,m^2,\ldots,m^L)$ where $m^i = (m^i_1,m^i_2,\ldots,m^i_k)$. Since $M$ is compressed, the bits of the message are such that each block takes on one of $2^k$ possible bit patterns uniformly at random.

Each $m^i$ is then channel coded using a nonsystematic LDPC code. This encoding is done in two steps. First, the $i$th $k$-bit message block $m^i$ is scrambled using a $k\times k$ matrix $S$ for $i=1,2,\ldots,L$. The scrambler $S$ is formed by generating random $k\times k$ binary matrices until one is found for which an inverse exists in GF(2). $S^{-1}$ is calculated using the LU factorization tailored to GF(2). The scrambling operation is $a^i = m^iS$. This yields a set of $L$ scrambled blocks $A$, each of size $k$. The scrambler is followed by a systematic $(N,k)$ LDPC encoding using the $k\times N$ generator matrix $G$ of a code $C$. The output codewords are then $b^i = a^iG$. There are $L$ of them, and each has length $N$.

The LDPC code in the encoder must meet certain criteria. It was shown in \cite{Harrison10_ITW} that a set of coded bits $R$ can be punctured from a codeword in $C$ such that the maximal stopping set in $R$ is the empty set, and the maximal stopping set in $R+v$ is nonempty $\forall v\notin R$. If a nonempty stopping set exists in the set of erasures in a given codeword, then there also exists a minimum-sized set of bits which must be guessed in order to allow the MP decoder to completely solve for the codeword. The size of this minimum-sized set is the number of degrees of freedom $D$. We require the code used by this encoder to have an associating $R$ such that $|R|=N-k$. The reason for such a choice is tied to a relationship between MP and maximum-likelihood (ML) decoders. It is shown in ~\cite[Lemma 3]{Harrison10_TIFS} that under such encoder constraints $D_{ML} = D_{MP} = |R_c|$, i.e. the number of degrees of freedom in the ML decoder and MP decoder are equal, and equivalent to the number of channel erased bits $|R_c|$. Thus, although the MP decoder is known to be suboptimal to the ML decoder, when $|R|=N-k$ the two provide the same performance. A random algorithm for finding such an $R$ was given in \cite{Harrison10_ITW}, and irregular degree distributions have been found which return puncturing patterns of this nature with reasonable probability \cite{Harrison10_TIFS}. Each codeword in the set of output codewords $B=(b^1,b^2,\ldots,b^L)$ is punctured according to $R$ to form the set of punctured codewords $P=(p^1,p^2,\ldots,p^L)$ where each $p^i=(p^i_1,p^i_2,\ldots,p^i_n)$. Since each punctured codeword has blocklength $n$, the effective \emph{rate} of this encoder is $k/n$.

Finally, these punctured codewords are distributed amongst $\eta$ packets in an interleaver denoted $\Pi$ so that $\alpha$ bits from all $L$ codewords are in each packet. The set of packets is denoted $X=(x^1,x^2,\ldots,x^\eta)$, and each packet $x^i$ has length $\alpha L$. Thus, $n = \eta\alpha$.\footnote{It is assumed that $\alpha$ divides $n$ for ease of notation.} Note that the interleaver $\Pi$ which spreads $\alpha$ bits from each punctured codeword into each packet, also yields equality in $D$ across codewords (see ~\cite[Corollary 1]{Harrison10_TIFS}).

\subsection{Decoder}\label{subsec:decoder}

The decoder consists of the following pieces: a buffer capable of holding all received data packets $Y$, a deinterleaving function $\Pi^{-1}$ which inverts the interleaving in the encoder to yield the punctured codewords $\tilde{P}$, the standard MP decoder with output codewords $\tilde{B}$ of which the systematic bits form the scrambled message blocks $\tilde{A}$, and finally the descrambler $S^{-1}$ which supplies an estimate of the compressed ciphertext $\tilde{M}$ as $\tilde{m}^i=a^iS^{-1}$. Notice that these collections of packets and codewords describe Bob's decoder. Eve's decoder is similar, except we denote the received data packets as $Z$, and then $\hat{P}$, $\hat{A}$, $\hat{B}$, and $\hat{M}$ form the eavesdroppers respective estimates of $P$, $A$, $B$, and $M$.

\section{Security for Correlated Channels}\label{sec:decodeAndSecrecy}
The security results for the correlated channel model from Section \ref{sec:modelAndCorrelation} are presented and compared with results given in \cite{Harrison10_TIFS} and \cite{Harrison10_ITW} for the independent channel case. We assume that the degree distribution for the LDPC code and the puncturing pattern $R$ are chosen such that $|R|=N-k$. Thus, the MP decoder achieves the ML performance, and furthermore, $D$ is equivalent for each decoded codeword.

\subsection{Main Security Results}

The first result is the distribution on $D$.
\begin{lemma}\label{lem:DisBinomial}
 Assume the encoder specified in Section \ref{subsec:encoder} where $|R|=N-k$. If $Q_m$ and $Q_w$ are memoryless, and erasures in the two channels are correlated, then the random variable $D$ which takes on the degrees of freedom in a received codeword is a scaled binomial random variable. Thus, for $1 \leq \beta \leq \alpha \eta$,
 \begin{equation}\label{eq:thm2}
   \Pr(D\geq\beta) = 1 - \sum_{i=0}^{\lceil\beta/\alpha\rceil - 1}{\eta \choose i}(1-\Pr(R_{ef}))^i \Pr(R_{ef})^{\eta-i}
 \end{equation}
 where $R_{ef}$ is the event that a particular packet is received error-free by Eve.
\end{lemma}
\begin{IEEEproof}
 The proof is identical to that for Lemma 4 in \cite{Harrison10_TIFS}, because although erasures in $Q_m$ and $Q_w$ are correlated, each packet is received error-free by Eve independent from other packets. This allows each transmitted packet to be considered a Bernoulli experiment as to whether Eve receives the packet. The sum of the missing packets is binomial with \emph{success} parameter $1-\Pr(R_{ef})$. The distribution in (\ref{eq:thm2}) follows.
\end{IEEEproof}
Another result which was found to apply to the independent erasure case in \cite{Harrison10_TIFS} follows directly from Lemma \ref{lem:DisBinomial} for correlated erasures as well.

\begin{theorem}\label{thm:expectedValD}
 If $|R|=N-k$ in the encoder, then $k/n = 1$, and
 \begin{equation}\label{eq:expectedD}
  \mathbb{E}[D] = H(X|Z) = (1-\Pr(R_{ef}))n = (1-\Pr(R_{ef}))k.
 \end{equation}
\end{theorem}
\begin{IEEEproof}
 Since $|R|=N-k$, then $n=|Q|=N-|R|=k$. Let us consider the model for a single codeword ($L=1$). We can then assume $\eta$ independent uses of a PEC with packets of length $\alpha$. Let $X=(x^1,x^2,\ldots,x^\eta)$ be the input to the channel, and $Z=(z^1,z^2,\ldots,z^\eta)$ be the output, where $\alpha$ bits are erased with probability $1-\Pr(R_{ef})$ or received error-free with probability $\Pr(R_{ef})$ with each channel use. The input distribution on $\alpha$ bits is uniform because the input distribution on $M$ is uniform, and the encoding function of rate one forms a bijection on $k$ bits. Thus, $H(x^i) = \alpha$ for $i=1,2,\ldots,\eta$. It can be shown that $H(z^i|x^i) = H(1-\Pr(R_{ef}))$, and $H(z^i) = H(1-\Pr(R_{ef})) + \Pr(R_{ef})\alpha$ (see \cite{Cover}, pg. 188). Then, $H(x^i|z^i) = H(z^i|x^i) - H(z^i) + H(x^i) = \alpha(1-\Pr(R_{ef}))$. Therefore, with $\eta$ independent uses of the channel (one for each packet), $H(X|Z) = (1-\Pr(R_{ef}))\eta\alpha = (1-\Pr(R_{ef}))n$. Since the mean of a binomial random variable is the product of its two parameters, $E[D/\alpha] = (1-\Pr(R_{ef}))\eta$, which concludes the proof.
\end{IEEEproof}

The remaining necessary task to characterize $D$ is to solve for $\Pr(R_{ef})$ when erasures in $Q_m$ and $Q_w$ are correlated with correlation coefficient $\rho$. If such an expression reverts back to the independent case when $\rho=0$, then as long as erasures are uncorrelated, and $Q_m$ and $Q_w$ are memoryless, the previous result for independent erasures applies. It was shown in \cite{Harrison10_ITW} that for statistically independent $Q_m$ and $Q_w$,
\begin{equation}\label{eq:independent_Ref}
   \Pr(R_{ef}) = \frac{1-\epsilon}{1-\epsilon\delta}.
\end{equation}

\begin{figure}
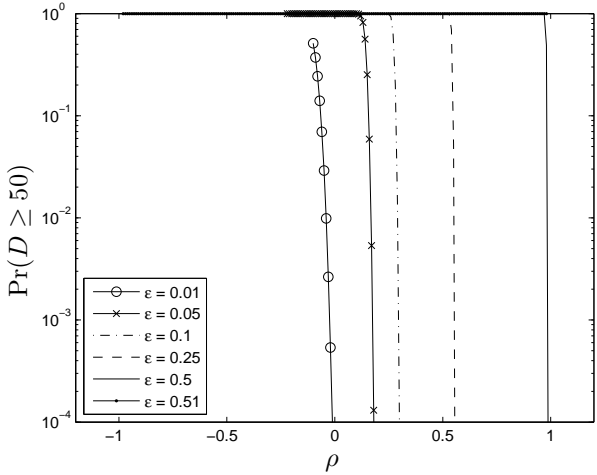

  \begin{lpic}{probDgeq50(0.6,0.6)}
    \lbl[l]{7,40,90;$\Pr(D\geq50)$}
    \lbl[t]{76,5;$\rho$}
  \end{lpic}
  \caption{$\Pr(D\geq 50)$ when the number of packets $\eta = 5000$, $\alpha = 1$, and Bob's erasure probability $\delta = 0.5$. Results are plotted for varying erasure probabilities $\epsilon$ for Eve's channel as a function of the correlation coefficient $\rho$.} \label{fig:probDgeq50}
\end{figure}

\begin{lemma}\label{lem:wiretap1}
 In the wiretap channel scenario with feedback, if channel erasures are correlated events across $Q_m$ and $Q_w$ with correlation coefficient $\rho$, then the probability that Eve obtains a single transmitted packet error-free is given as
 \begin{equation}\label{eq:correlated_Ref}
   \Pr(R_{ef}) = \frac{1-\epsilon}{1-\epsilon\delta-\rho\sqrt{\delta\epsilon(1-\delta)(1-\epsilon)}}.
 \end{equation}
\end{lemma}
\begin{IEEEproof}
Let $W$ be the total number of times that Bob requests a single packet over $Q_m$ before he obtains it error-free. Since $Q_m$ is memoryless, the packet is erased each time independently with probability $\delta$. Therefore, $W$ is a random variable which takes on the number of total transmissions up to and including the first successful reception of the packet, and is thus geometrically distributed with success parameter $1-\delta$. Then, $\Pr{(W=j)}=(1-\delta)\delta^{j-1}$ \cite{Grimmett_ProbBook}. Let $E_m^i$ and $E_w^i$ denote the respective erasure outcomes in $Q_m$ and $Q_w$ for the $i$th retransmission of the packet, where a one signifies an erased packet as before. Therefore,
\begin{eqnarray}\label{eq:prRef}
    \Pr{(R_{ef})} &=& \sum_{j=1}^\infty\Pr(R_{ef}|W=j)\Pr(W=j) \nonumber \\
    &=& \sum_{j=1}^\infty(1-\Pr(E_w^1=\cdots=E_w^j=1|E_m^1=\cdots \nonumber \\
    && =E_m^{j-1}=1,E_m^j=0)\Pr(W=j) \nonumber \\
    &=& \sum_{j=1}^\infty \left(1 - \left(\frac{p_{11}}{\delta}\right)^{j-1}\frac{p_{01}}{1-\delta}\right)(1-\delta)\delta^{j-1} \nonumber \\
    &=& (1-\delta)\sum_{j=1}^\infty\delta^{j-1}-p_{01}p_{11}^{j-1} \nonumber \\
    &=& 1 - \frac{p_{01}}{p_{11}}\left[\left(\sum_{j=0}^\infty p_{11}^j\right) - 1\right] \nonumber \\
    &=& \frac{1-p_{11}-p_{01}}{1-p_{11}} \nonumber.
\end{eqnarray}
Now since $p_{11} = \rho\sqrt{\delta\epsilon(1-\delta)(1-\epsilon)} + \delta\epsilon$ and $p_{01} = \rho\sqrt{\delta\epsilon(1-\delta)(1-\epsilon)} + \epsilon(1-\delta),$ then
\begin{displaymath}
 \Pr(R_{ef}) = \frac{1-\epsilon}{1-\epsilon\delta-\rho\sqrt{\delta\epsilon(1-\delta)(1-\epsilon)}}.
\end{displaymath}
\end{IEEEproof}
Clearly, this reduces to the independent case in (\ref{eq:independent_Ref}) when $\rho=0$. Fig. \ref{fig:probDgeq50} gives an example of $\Pr(D\geq\beta)$ using this expression for $\Pr(R_{ef})$ where $\beta$ is chosen to be 50. The figure assumes $C$ to be rate-$1/2$ with blocklength $N = 10000$, $|R|=N-k=5000$, and $\alpha = 1$. Therefore, $\eta=n/\alpha = 5000$. We set $\delta=0.5$ and plot different $\epsilon$ values as $\rho$ takes on all possible values indicated by the bounds in (\ref{eq:rhoBounds}). Fig. \ref{fig:probDgeq50} implies the existence of a correlation threshold $\rho_{th}$ for $\Pr(D\geq\beta)$, in that if all other parameters are set, then for $\rho < \rho_{th}$, $\Pr(D\geq\beta)\approx 1$. Yet, for $\rho > \rho_{th}$, $\Pr(D<\beta)$ is essentially one. Notice, however, that when $\epsilon = 0.51$, $\Pr(D\geq50)\approx 1 \forall \rho$.

\subsection{Security at the Correlation Bounds}

Now that we understand how correlation affects the random variable $D$, we evaluate the extreme cases in correlation coefficients by considering the bounds on $\rho$ in (\ref{eq:rhoBounds}).

Consider the lower bound $\rho = \frac{\max(\delta+\epsilon-1,0)-\delta\epsilon}{\sqrt{\delta\epsilon(1-\delta)(1-\epsilon)}}$. Then,
\begin{eqnarray}
 \Pr(R_{ef}) = \frac{1-\epsilon}{1-\max(\delta+\epsilon-1,0)}.
\end{eqnarray}
Therefore,
\begin{equation}
\Pr(R_{ef}) = \left\{
\begin{array}{ll}
 \frac{1-\epsilon}{2-\delta-\epsilon} & \mbox{if $\delta + \epsilon > 1$} \\ 1-\epsilon & \mbox{otherwise}
\end{array}
\right.
\end{equation}
If $\delta +\epsilon > 1$, this implies that $\Pr(R_{ef}) > 1-\epsilon$, which of course is worse for secrecy than if $\delta + \epsilon\leq 1$, all other things being equal. When $\delta + \epsilon\leq 1$, $\Pr(R_{ef})=1-\epsilon$ implies that negative correlation can reduce the eavesdropper to an effective erasure channel where only one chance is given to intercept each packet, despite retransmission of some packets. Of course, the reasoning behind this is that this minimum correlation indicates that all of Eve's missing packets are obtained by Bob in the first transmission with probability one.

Now consider the upper bound $\rho = \frac{\min(\delta,\epsilon)-\delta\epsilon}{\sqrt{\delta\epsilon(1-\delta)(1-\epsilon)}}$.
Then,
\begin{equation}
\Pr(R_{ef}) = \frac{1-\epsilon}{1-\min(\delta,\epsilon)},
\end{equation}
which then implies that when Eve has at least as good of a channel as Bob, i.e. $\delta \geq \epsilon$, that the upper bound yields perfect correlation, that is every packet is eventually received by Eve error-free. However, if Bob can maintain a channel advantage over Eve, i.e. $\delta < \epsilon$, then we see that $\Pr(R_{ef}) = \frac{1-\epsilon}{1-\delta} < 1$. Thus, even maximum correlation cannot reduce $\mathbb{E}[D]$ to zero. Since $\mathbb{E}[D]$ grows with $k$ in (\ref{eq:expectedD}), this indicates that we can still gain as many degrees of freedom as we desire on average by increasing the dimension of the encoder.

\section{Conclusions}\label{sec:conclusion}
In conclusion, we have analyzed the secrecy coding and ARQ scheme from \cite{Harrison10_ITW} for correlated packet erasures in the main and wiretap channels. Security results have been compared with the results previously known for independent erasures in the two channels. Furthermore, the overall secrecy effect of correlation is maximized when $\rho$ takes on its minimum value resulting in only one opportunity for an eavesdropper to obtain every packet. The minimum security due to correlation occurs at the upper bound of the correlation coefficient, and can reduce the degrees of freedom to zero, although if Bob can maintain even the slightest advantage in overall channel quality, the degrees of freedom cannot go to zero, and can be set arbitrarily by changing the dimension of the encoder $k$. Thus, the physical-layer coding scheme retains its security enhancement features in spite of correlation across erasure channels.

\bibliographystyle{ieeetran}
\bibliography{references_correlation}

\end{document}